# Detecting Requirements Defects Utilizing A Mathematical Framework for Behavior Engineering

*Kushal Ahmed, Toby Myers, Lian Wen and Abdul Sattar

Institute for Integrated and Intelligent Systems
Griffith University
Nathan, QLD 4111, Australia
{k.ahmed, t.myers, l.wen, a.sattar}@griffith.edu.au

*Abstract*— Behavior Engineering (BE) provides a rigorous way to derive a formal specification of a software system from the requirements written in natural language. Its graphical specification language, Behavior Tree (BT), has been used with success in industry to systematically translate large, complex, and often erroneous requirements into an integrated model of the software system. BE's process, the Behavior Modeling Process (BMP), allows requirements to be translated into individual requirement BTs one at a time, which are then integrated to form a holistic view of the system. The integrated BT then goes through a series of modifications to construct a specification BT, which is used for validation and verification. The BMP also addresses different types of defects in the requirements throughout its process. However, BT itself is a graphical modeling notation, and the types of integration relations, how they correspond to particular issues, how they should be integrated and how to get formal specification have not been clearly defined. As a result, the BMP is informal, and provides guidelines to perform all these tasks on an ad-hoc basis. In this paper, we first introduce a mathematical framework which defines the graphical form of BTs which we use to define the integration relationships of BTs and to formalize the integration strategy of the BMP. We then formulate semi- automated requirements defects detection techniques by utilizing this underlying mathematical framework, which may be extended to formalize the BMP, develop change management framework for it, build techniques for round-trip engineering and so on.

## I. INTRODUCTION

The complexity of software intensive systems is ever-increasing, which is creating larger and more complex specifications. Defining a formal specification using graphical modeling notations allows further formal analysis (e.g. validation and verification) to be performed prior to development. One of the difficult parts of developing systems is to derive a formal specification from a Software Requirements Specification (SRS). But, an SRS is commonly written in Natural Language (NL) which may often be erroneous and ill-defined. As a result, the transition from an SRS into a formal specification is further complicated due to the presence of these defects.

Behavior Engineering (BE) [1] is a scalable and repeatable end-to-end methodology to model the NL requirements of a software intensive system. The BMP captures the behaviors of a software system into the integrated views of a Behavior Tree (BT) and a Composition Tree (CT). Whilst modeling the requirements, BE also provides a rigorous way to address the imperfect knowledge in the requirements. The effectiveness and efficiency of BE to perform this task has been demonstrated through modeling large-scale industrial software systems [2].

Many different defects detection techniques have been proposed that utilise UML diagrams ([3]–[5]). However, UML diagrams represent variety of aspects of the system across several diagrams, which could lead to inconsistent interpretations. Having different syntax and semantics, UML diagrams are difficult to combine to derive an integrated view of the software system. The multiple partial views of the system, which have a degree of overlap, along with the contextual switch when moving between diagrams make it hard to detect many types of defects, particularly those that involve interactions between the requirements [6].

In contrast, a key characteristic of BE is that the same modeling notations are used throughout the modeling process. BTs preserve the original intention of the requirements in a structured easy-to-understand visual representation. They essentially capture partial interleaved state machines of the components of the software system. As a result, it becomes easy to combine individual requirements to create a formal model from the requirements without any miraculous leap, which in turn promotes requirements defects detection earlier in the software life-cycle.

Moreover, BE tames the complexity of the requirements by translating each requirement independently into its own BT. Formalising BTs helps the requirements analysts to increase their own understanding about the scenarios contained in the requirements. Thus it aids detecting ambiguous, inconsistent or missing requirements. After the translation stage, BTs are integrated to form a holistic view of the system. At this stage, the integration relations formed by BTs are the central means to detect defects in the requirements. Once the BTs are integrated into one BT, further modifications reveal many other types of defects (e.g. *inconsistent*, *incomplete*, *incorrect* etc.) in the requirements.

However, BTs are graphical modeling notation. The BMP only provides an informal guideline to address the imperfect knowledge in the requirements and produce a model of the software system. As a result, how BTs form integration relations, how different types of integration relations correspond to particular issues, how they should be integrated and how to get formal specification have not themselves been formally defined. That is, BE lacks in an underlying framework that formalises all these tasks of the BMP.

In this paper, we first provide a mathematical framework that defines the graphical form of BTs which we use to define different types of integration relations in BTs. This framework formalises the integration strategy of the BMP which is an important part of the integration process of the BMP. Then, we demonstrate one of the benefits of such framework by deriving semi-automated techniques for detecting defects in the requirements. In particular, we formulate mappings of some special types of integration relations to incomplete, ambiguous, incorrect and redundant defects as per IEEE Std 830 requirements characteristics [7].

This paper is structured as follows. Section II provides the background about BE. Section III presents a mathematical behavioral model abstracting BT. Section IV formalises the integration relations of the behavioral models including special cases of these relations. Section V formulates the detection of requirements defects. Section VI illustrates the formal integration relations and defects detection by a case study, and evaluates our proposed approach. In Section VII, we describe the related work. Finally, Section VIII mentions the future directions and concludes the paper.

## II. BACKGROUND

In this section, we provide a brief introduction to Behavior Engineering (BE), a combination of scenario-based and state-based requirements engineering approach ([1], [8]). Traditional system modeling approaches build a model and then try to satisfy a set of requirements. In contrast, BE starts with translating the NL requirements into BTs one at a time and then integrating them all to produce an integrated model of the system. The BMP uses the Behavior Modeling Language (BML) that includes Behavior Tree (BT) and Composition Tree (CT)[1]. In this section, we briefly describe the constructs of BT, and how the BMP works.

---
[1]Composition Tree [8, p-66] is out of the scope of this paper.

### A. Behavior Tree

A BT is a tree-based graphical form that represents the behavior of individual or networks of entities which realise and change states, create and break relations, make decisions, respond to and cause events, and interact by exchanging information and passing control [9]. The tree structure of the BT is formed by connecting the nodes by the edges, where each node captures a unit of behavior associated with a component (entity) and each edge indicates a flow of control.

To capture a unit of behavior, a node in BT stores a set of attributes (see Fig. 1):

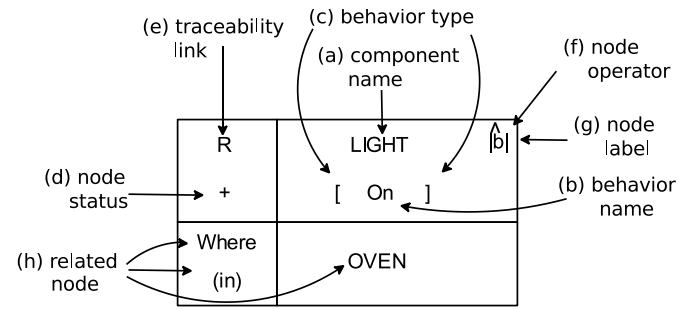

Fig. 1. Attributes that a BT node may hold

(a) *Component name:* The name of the component (e.g. Light).

(b) *Behavior name:* The name of the behavior associated with the component (e.g. behavior 'On' of Light).

(c) *Behavior type:* The type of the behavior of the component. The types may be state realisation, selection, event, guard and internal/external input/output. These different types support modeling control flows with deterministic or nondeterministic choice.

(d) *Node status:* The status of the node indicating the level of modifications made apart from the original NL requirement. A node can have status like original, implied(+), missing(-), design(+-), updated(++) and deleted(--). (e) Traceability link: The link to the unique ID of one or more original NL requirements.

(f) *Node operator:* To support alternative, circular or concurrent behaviors, a node may indicate different operators like synchronisation(=), reversion(^), reference(=>), branch kill(--), may(%), conjunction(&), disjunction(j) and exclusive OR (XOR).

(g) *Node label:* Labels may be needed for disambiguation if there exist multiple destinations for an operator.

(h) *Related nodes:* A behavior may associate another component forming a relation. A relation may be attributed by the component name, its qualifier (what, where etc.) and preposition (in, on etc.). For example, the Oven is related to the Light since the behavior 'On' of Light happens in Oven (Where(in)).

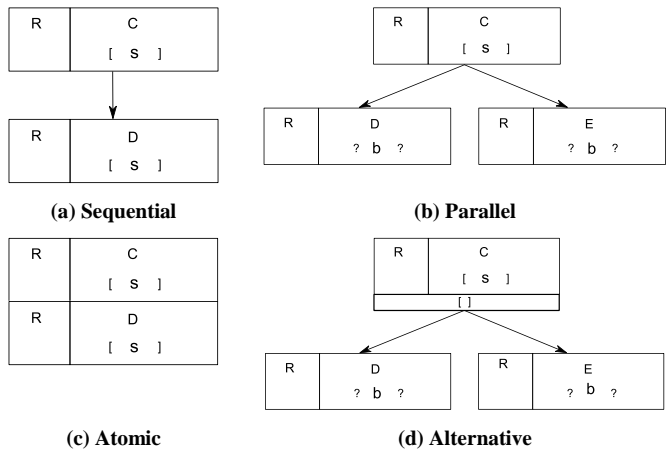

**Fig. 2. A BT may have (a) sequential flow of control, (b) multiple flows, (c) atomic flow, and (d) single sequential flow selected from many flows.**

BTs have four types of connecting edges to depict the flow of control (see Fig. 2):

(a) *Sequential:* Sequential flow of control, where the behavior of concurrent nodes may be interleaved between two nodes.

(b) *Parallel:* Control is passed to all the child nodes.

(c) *Atomic:* Sequential flow of control, but no interleaved behavior is allowed.

(d) *Alternative:* Allows passing the control only to one child node out of several nodes.

A BT manages cyclic, concurrent or synchronised behaviors by using different operators that link the BT nodes implicitly, which keeps the structure simple to view the flow of control and concurrency in the model. Thus BT does not overwhelm a reader with a web of complex interconnections that would be created by a network structure even if it stores large amount of complex information. The detailed description of BT notation can be found at ([8, ch-3], [10]) and the semantics at [11].

*B. Behavior Modeling Process*

The first stage of the BMP is to translate the NL requirements by representing them in BTs one at a time. The BTs originated from the NL requirements are referred as Requirement Behavior Trees (RBTs). Fig. 3 shows the BTs for the requirements R1 (right) and R6 (left) of the Micro-wave oven (see Table III at Appendix A). In Fig. 3 (right), the first node models the condition that the Door has to be in the state Closed. The second node specifies a guard; the control flow cannot proceed until the guard is satisfied, and so on. The requirements analyst gains more understanding about the system whilst all behavioral scenarios are depicted in RBTs, and many different types of defects like ambiguity, inconsistent naming of the components or entities, missing behavior or absence of cause and effect are identified during this translation.

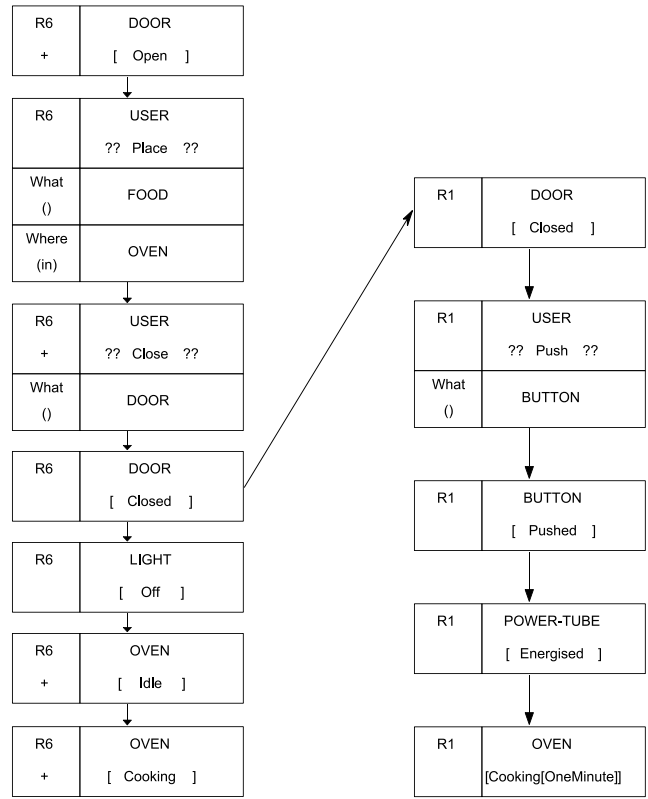

**Fig. 3. A branch node of (a) RBT6 matches with the root node of (b) RBT1, hence they form an integration relation where RBT6 is the parent and RBT1 is the child of the relation.**

Once all the BTs are formed, the BMP finds the common nodes in BTs for integration. For example, a branch node of RBT6 matches with the root node of RBT1, which causes an integration relation to be formed between them (the arrow in Fig. 3 shows this relation). It implies that they can be integrated to form one BT. In this requirements integration stage, all the BTs are integrated to form an Integrated Behavior Tree (IBT). If any RBT cannot be integrated, it indicates defects in that RBT and its associated requirements. The BMP can then be followed to further process the IBT to derive an executable model of the system, which can be simulated for validation [12] and model-checked for verification ([13]–[17]).

III. GENERIC BEHAVIORAL MODEL

In order to formalise integration relations in BTs, we define a generic behavioral model using mathematical notation. Fig. 4(a) shows a requirement BT for the requirement R4 of micro-wave oven system (see Table III at Appendix A) having two nodes and one edge. Each node represents a behavioral unit associated with the component, and the edge shows the sequential flow of control.

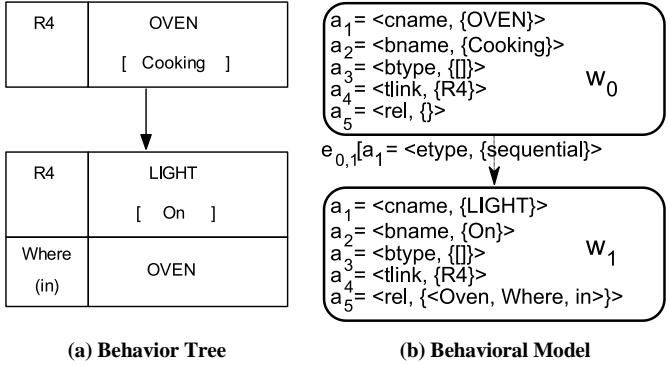

(a) Behavior Tree       (b) Behavioral Model

**Fig. 4. Representation of (a) a BT by (b) a generic Behavioral Model**

Here, each node (behavioral unit)[2] represents a unit of behavior of the corresponding entity and that behavior is expressed by a set of attributes. For example, the root node has the component name (*cname*) of OVEN, the behavior name (*bname*) of Cooking, the behavior type (*btype*) of state realisation ([]), the traceability link (*tlink*) of R4 and so on. Again, the edge shows the flow of control and has the type of sequential. Fig. 4(b) shows a sample mathematical representation of the BT shown at Fig. 4(a). Additionally, each edge might have other attributes (e.g. time constraints). Therefore, both a *behavioral unit* (node) and an edge in BT consist of a set of attributes where each attribute has a *name* and *value* associated with it. We formally define attribute, behavioral unit and edge below:

*Definition 1:* An **Attribute**, denoted by a, consists of a pair denoted by <*m, v*> where *m* is the name and *v* is the value of the attribute.

The value of an attribute can be of any type (e.g. string, number, boolean etc.). For example, the value of the component name (*cname*) attribute is a string (*OVEN*). The value can be a set as well. For example, the value of the traceability link (*tlink*) can be {R1, R4}, where 'R1' and 'R4' represent the identities of two requirements.

*Definition 2:* A **Behavioral Unit** (BU), denoted by w, represents a unit of behavior of individual or networks of entities, and consists of a set of attributes $\{a_1, a_2, ..., a_{n_w}\}$. Here, $n_w$ is the number of attributes that a BU holds.

In Fig. 4(b), the root BU represents a unit of behavior of *OVEN* and expressed by a set of attributes: $a_1$=<*cname*, {*OVEN*}>, $a_2$=<*bname*, {*Cooking*}> and so on.

*Definition 3:* An **Edge**, denoted by e, represents the flow of control of BUs, and consists of a set of attributes $\{a_1, a_2, ... a_{n_e}\}$. Here, $n_e$ is the number of attributes that an edge holds.

In Fig. 4(b), the two BUs $w_0$ and $w_1$ are connected by the edge $e_{0,1}$ that has one attribute $a_1$=<*etype*, {*sequential*}>. Now, we define the generic behavioral model as a tree consisting of BUs, edges and their relations.

*Definition 4:* The **Behavioral Model** (BM) of a software system, denoted by *b*, is a tree expressed in a tuple < *W, E, Γ* > where

- *W* is a tuple denoted by < $r_b$, $N_b$, $L_b$ > where $r_b$ is the root BU, $N_b$ is the set of intermediate (non-root and non-leaf) BUs and $L_b$ is the set of leaf BUs.
- *E* is the set of edges for connecting the BUs in *W*.
- *Γ* is the set of triples ($w_i$, $e_{i,j}$, $w_j$) where $w_i \in \{r_b \cup N_b\}$, $w_j \in \{N_b \cup L_b\}$ and $e_{i-j}$ is an edge from $w_i$ to $w_j$ i.e. $w_i$ is the parent of $w_j$, $w_i \neq w_j$.

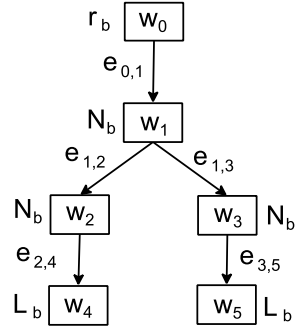

**Fig. 5. Generic Representation of A Behavioral Model that has root, branch and leaf Behavioral Units connected by Edges in a tree structure.**

Fig. 5 is an example representation of a generic BM. Here, $r_b = w_0$, $N_b = \{w_1, w_2, w_3\}$, $L_b = \{w_4, w_5\}$, $E = \{e_{0,1}, e_{1,2}, e_{1,3}, e_{2,4}, e_{3,5}\}$ and $Γ = \{(w_0, e_{0,1}, w_1), (w_1, e_{1,2}, w_2), (w_1, e_{1,3}, w_3), (w_2, e_{2,4}, w_4), (w_3, e_{3,5}, w_5)\}$.

IV. FORMAL RELATIONS OF BEHAVIORAL MODELS

According to the precondition and interaction axioms of BE [1], each BM has a precondition (root BU) that needs to be satisfied in order for the BUs encapsulated in the BM to be applicable, and that precondition must be established by at least another BM. Therefore, the BMs are interrelated for integration. In this section, we formalise these integration relations extending these axioms.

If the root BU of a BM is equivalent to any BU in another BM, they can be integrated to form one BM. The relation of such type is called primary integration relation. Since a BU consists of a set of attributes, the equivalency of two candidate BUs depends on the attributes they hold. Hence, we first define the similarity of two attributes. Then, we formulate the equivalency of two BUs. Finally, we formalise the integration relations in BMs.

*A. Similarity in Attributes*

Two attributes, given their names are equal (or have the same mapping), can be similar even if the values are not equal. So we consider similarity between two attributes as a fraction value ranging from 0 to 1.

---

[2]We will use behavioral unit and node terms interchangeably depending on the context.

Definition 5: The *Similarity* measure of the two attributes $a_1$ and $a_2$, denoted by $a_1 \odot a_2$, is defined by:

$$a_1 \odot a_2 = \begin{cases} 0 & \text{if } a_1.m = a_2.m, \text{ or} \\ & \text{atrmap}(a_1) = \text{atrmap}(a_2) \\ \xi(a_1, a_2) & \text{otherwise} \end{cases}$$

where $0 \leq \xi(a_1, a_2) \leq 1$. Here $\xi$ is a similarity function that calculates the similarity between the values of the attributes given their names are equal or have same mapping i.e. $\text{atrmap}(a_1) = \text{atrmap}(a_2)$.

The similarity function $\xi$ can be defined in many ways depending on the domain information of the software system. In its simplest form, it can return 1 if the values of the attributes are equal and 0 otherwise. For example,

$$\xi(a_1, a_2) = \begin{cases} 1 & \text{if } a_1.v = a_2.v \\ 0 & \text{otherwise} \end{cases}$$

In more complex cases, the attributes can be *compatible* with each other for integration even if the values differ. For example, two candidate BT nodes can have attribute named *btype* (behavior type), and the value for one node can be *state realisation* ([]) and another can be *selection* (?). In this case, these values are *compatible* with each other for integration. Alternatively, the values *state realisation* ([]) and *internal input* (><) of *btype* attribute are *incompatible* for integration. Therefore, the similarity function $\xi$ can return a fraction $\beta$ ($0 < \beta < 1$) if the values are compatible. That is,

$$\xi(a_1, a_2) = \begin{cases} 1 & \text{if } a_1.v = a_2.v \\ \beta & \text{else if } a_1.v \text{ and } a_2.v \text{ are compatible} \\ 0 & \text{otherwise} \end{cases}$$

where $\beta$ is a similarity fraction ($0 < \beta < 1$).

However, the requirements analyst has to decide which pair of values of an attribute is *compatible* or which one is not. For storing this *compatibility* information, a two-dimensional matrix can be built from the set of all allowed values of an attribute.

*Definition 6:* The value $v$ of an attribute $a$ are taken from a **set of allowed values** denoted by $V_a = \{v_1, v_2, \ldots, v_{n_a}\}$. Here, $n_a$ is the size of that set.

For example, the behavior type btype attribute may have *state realisation* ([]), *selection* (?), *event* (??), *guard* (???), *internal input* (><), *internal output* (<>), *external input* (>><<) and *external output* (<<>>) values. So, these values can be taken as the elements of the set of allowed values $V_a$ for *btype* attribute.

*Definition 7:* The **compatibility matrix** of an attribute, denoted by $M_a$, is a two-dimensional upper-half square matrix of size $(n_a \times n_a)/2$ and defined by:

$$M_a[i,j] = \begin{cases} 1 & \text{if } v_i, v_j \in V_a \text{ compatible} \\ 0 & \text{otherwise} \end{cases}$$

where, $i = 1$ to $n_a$, $j = i + 1$ to $n_a$, $i \neq j$.

In some cases, an attribute can have a set of values. We may consider the values to be *similar* if they hold a subset or superset relationship. In other case, if they have common values, we may consider percentage of matched values. Otherwise, we can assume a *similarity fraction $\beta$* if all values are *compatible* to each other. In this case, we may define the similarity function $\xi$ by the following way:

$$\xi(a_1, a_2) = \begin{cases} 1 & \text{if } a_1.v \subseteq a_2.v \text{ or } a_1.v \supseteq a_2.v \\ \dfrac{|a_1.v \cap a_2.v|}{|a_1.v \cup a_2.v|} & \text{else if } a_1.v \cap a_2.v \neq \Phi \\ \beta & \text{else if } a_1.v \text{ and } a_2.v \text{ are compatible} \\ 0 & \text{otherwise} \end{cases}$$

In typical cases, we can assume the similarity fraction $\beta = 0.50$ if the candidate values are compatible. The appropriate similarity measure can be even more complex depending on the integration strategies.

*B. Equivalency in Behavioral Units*

To determine the *equivalency* of two BUs, we use the *similarity* measure of the each of the attributes to determine their *similarity*. If the *similarity* is greater or equal to a threshold value, we consider this BUs as *equivalent*.

All the attributes that form a BU may not have the same significance when we calculate the similarity. For example, the *component name* (*cname*) attribute has a higher significance than that of the *behavior type* (*btype*). Again, the *traceability link* (*tlink*) has a very low or no significance compared to *cname* and *btype*. Additionally, the attributes in one BU might evolve in time as well. Therefore, we consider a weight value as a measure of significance of each of the attributes.

*Definition 8:* Each attribute $a_i \in w$, where $i = 1$ to $n_w$, has a **Weight** value, denoted by $g_{a_i}$, which determines the significance of that attribute.

Now, we calculate the similarity between two candidate BUs:

*Definition 9:* For two BUs, $w_1 = \{a_1, a_2, \ldots, a_{n_w}\}$ and $w_2 = \{a_1, a_2, \ldots, a_{n_w}\}$, such that for each $a_i \in w_1$ and $a_j \in w_2$, $i = j = 1$ to $n_w$, $a_i.m = a_j.m$ or $\text{atrmap}(a_i) = \text{atrmap}(a_j)$, their **Similarity** measure, denoted by $w_1 \odot w_2$, is defined by:

$$w_1 \odot w_2 = \frac{\sum_{k=1}^{n_w}(w_1.a_k \odot w_2.a_k) * g_{a_k}}{\sum_{k=1}^{n_w} g_{a_k}}$$

*Similarity* of two BUs $w_1 \odot w_2$ is a probabilistic measure, which we use to define the *equivalent* BUs:

*Definition 10:* Two BUs $w_1$ and $w_2$ are **Equivalent**, denoted by $E(w_1, w_2, \alpha) = 1$, if $w_1 \odot w_2 \geq \alpha$, where $0 \leq \alpha \leq 1$. That is,

$$E(w_1, w_2, \alpha) = \begin{cases} 1 & \text{for } w_1 \odot w_2 \geq \alpha \\ 0 & \text{otherwise} \end{cases}$$

where $\alpha$ is equivalency threshold value.

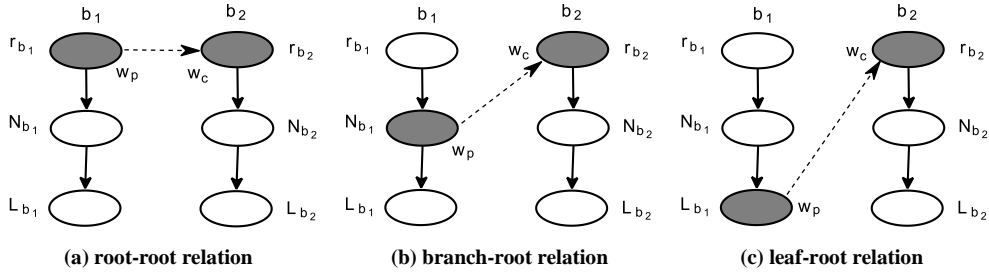

**(a) root-root relation**  **(b) branch-root relation**  **(c) leaf-root relation**

**Fig. 6.** *Primary integration relations:* **(a)** The root BUs of two BMs may be equivalent, **(b)** A branch BU may be equivalent to a root BU, and **(c)** A leaf BU may be equivalent to a root BU.

Let's assume that each BU has six attributes that have names *component name* $a_1.\mathrm{m} = cname$, *behavior name* $a_2.\mathrm{m} = bname$, *behavior type* $a_3.\mathrm{m} = btype$, *traceability link* $a_4.\mathrm{m} = tlink$, *node status* $a_5.\mathrm{m} = status$ and *related node* $a_6.\mathrm{m} = rel$. For determining the *equivalent* BUs, the weights of *cname* $g_{a1} = 50\%$, *bname* $g_{a2} = 35\%$, *btype* $g_{a3} = 15\%$ and all the rest as 0%, and the *equivalency threshold* α = 1. With this settings, Fig. 3 shows that the root of RBT1 ($w_1$) has an integration relation with an intermediate BU of RBT6 ($w_2$). Here, $w_1 = \{a_1 =<cname, DOOR >, a_2 =<bname, Closed>, a_3=<btype, []>, a_4 =<tlink, R1>, a_5=<status, (original)>; a_6=<rel, \{\}>\}$, and $w_2 = \{a_1=<cname, DOOR>, a_2=<bname, Closed>, a_4=<btype, []>, a_4=<tlink, R6>, a_5=<status, (original)>, a_6=<rel, \{\}>\}$. Therefore, $w_1$ and $w_2$ are *equivalent* for integration (Definition 10).

The weight values $g_{a1}, g_{a2}, g_{a3}, \ldots, g_{a_{n_w}}$, of the attributes, along with *equivalency threshold* α, *similarity fraction* β and *similarity function* ξ, depend on the integration strategy. Much more complex combinations of these parameters are possible for determining the *equivalent* BUs, but are beyond the scope of this paper.

*C. Primary Integration Relations in Behavioral Models*

When equivalent BUs exist in multiple BMs, they may establish integration relations. However, since the validity of integration relations depends on the overall context of the specification, all possible/identified relations may not be appropriate. Therefore, the requirements analyst has to decide which integration relations to accept or reject. If the root BU of one BM is *equivalent* to a BU in another BM, they form a *primary integration relation* which we define by the following way:

*Definition 11:* Primary Integration Relation: A BM $b_1 =<W_1, E_1, \Gamma_1>$ forms a *primary integration relation* with another BM $b_2 =<W_2, E_2, \Gamma_2>$ iff $\exists w_p \in \{r_{b1} \cup N_{b1} \cup L_{b1}\}$ and $w_c = r_{b2}$ such that for each $(w_p, w_c)$, $E(w_p, w_c, α) = 1$. We denote it by $R(b_1, b_2)$, which is a set of *equivalent* BUs pairs i.e. $R(b_1, b_2) = \{(w_p, w_c): w_p \in \{r_{b1} \cup N_{b1} \cup L_{b1}\}, w_c = r_{b2}, E(w_p, w_c, α) = 1\}$.

Here, we refer $b_1$ as the *parent* BM and $b_2$ as the *child* BM of the relation $R(b_1, b_2)$. Similarly, $w_p$ is referred as the *parent* integration BU and $w_c$ as the *child* integration BU. We identified three types of *primary integration relations*: *root-root*, *branch-root*, *leaf-root*.

*1) root-root* relation: If the root BU of a BM is *equivalent* to the root BU of another BM, they establish a *root-root* relation (Fig. 6(a)).

*Definition 12: root-root* relation: A BM $b_1$ has a *root-root* relation with another BM $b_2$ if $w_p = r_{b1}$, $w_c = r_{b2}$ and $E(w_p, w_c, α) = 1$. We denote it by: $R_{\text{r-r}}(b_1, b_2) = \{(w_p, w_c): w_p = r_{b1}, w_c = r_{b2}, E(w_p, w_c, α) = 1\}$.

*2) branch-root* relation: If an intermediate (branch) BU of a BM is *equivalent* to the root BU of another BM, they establish a *branch-root* relation (Fig. 6(b)).

*Definition 13: branch-root* relation: A BM $b_1$ forms a *branch-root* relation with another BM $b_2$ if $\exists w_p \in N_{b1}$ and $w_c = r_{b2}$ such that for each $(w_p, w_c)$, $E(w_p, w_c, α) = 1$. We denote it by $R_{b\text{-r}}(b_1, b_2) = \{(w_p, w_c): w_p \in N_{b1}, w_c = r_{b2}, E(w_p, w_c, α) = 1\}$.

*3) leaf-root* relation: If a leaf BU of a BM is *equivalent* to the root BU of another BM, they establish a *leaf-root* relation (Fig. 6(c)).

*Definition 14: leaf-root* relation: A BM $b_1$ forms a *leaf-root* relation with another BM $b_2$ if $\exists w_p \in L_{b1}$ and $w_c = r_{b2}$ such that for each $(w_p, w_c)$, $E(w_p, w_c, α) = 1$. We denote it by $R_{l\text{-r}}(b_1, b_2) = \{(w_p, w_c): w_p \in L_{b1}, w_c = r_{b2}, E(w_p, w_c, α) = 1\}$.

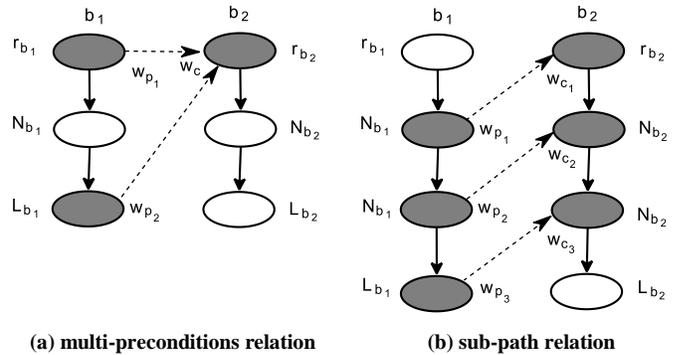

**(a) multi-preconditions relation**  **(b) sub-path relation**

**Fig. 7.** *Special integration relations:* **(a)** Multiple BUs may be equivalent to the root BU, and **(b)** A sub-path of BUs may be equivalent to a sub-path BUs.

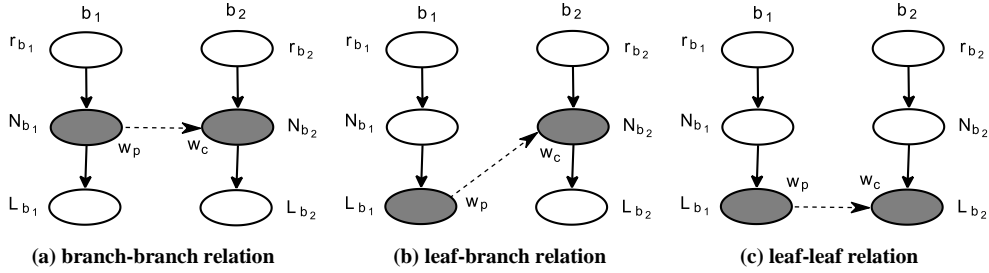

(a) branch-branch relation  (b) leaf-branch relation  (c) leaf-leaf relation

**Fig. 8.** *Special integration relations (non-root):* **(a)** A branch BU may be equivalent to a branch BU, **(b)** A leaf BU may be equivalent to a branch BU, and **(c)** A leaf BU may be equivalent to a leaf BU.

*D. Special Integration Relations*

In addition to the integration relations defined previously, there also exist several special cases. BMs may have multiple integration relations. They may be missing preconditions, and integration may be possible by allowing interleaving of different preconditions. We identified three special integration relations: *multi-preconditions relation*, *sub-path relation* and *non-root relation*. We formalise each type of relations below:

*1) multi-preconditions relation*: Two BMs form multi-preconditions relation if the root BU of one BM is equivalent to multiple BUs in another BM (Fig. 7(a)).

*Definition 15: multi-preconditions* relation: A BM $b_1$ has a *multi-preconditions* relation with another BM $b_2$, denoted by $R_{m-p}(b_1, b_2) = 1$, iff $\exists (w_{p1}, w_{c1}) \in R(b_1, b_2)$, $\exists (w_{p2}, w_{c2}) \in R(b_1, b_2)$ such that $w_{p1} \neq w_{p2}$ and $w_{c1} = w_{c2}$.

*2) sub-path relation*: Two BMs form *sub-path* relation if the BUs of a sub-path of one BM are *equivalent* to the same order of a sub-path of another BM. We consider at least three consecutive *equivalent* BUs to form a sub-path relation (Fig. 7(b)).

*Definition 16: sub-path* relation: A BM $b_1$ has a *sub-path* relation with another BM $b_2$, denoted by $R_{s-p}(b_1, b_2) = 1$, iff $\exists \{w_1, w_2, w_3, \ldots, w_s\} \subseteq \{r_{b1} \cup N_{b1} \cup L_{b1}\}$ where $\{(w_1, e_{1-2}, w_2), (w_2, e_{2-3}, w_3), \ldots, (w_{(s-1)}, e_{(s-1)-s}, w_s)\} \subseteq \Gamma_{b1}$ and $\exists \{w_1, w_2, w_3, \ldots, w_t\} \subseteq \{r_{b2} \cup N_{b2} \cup L_{b2}\}$ where $\{(w_1, e_{1-2}, w_2), (w_2, e_{2-3}, w_3), \ldots, (w_{(t-1)}, e_{(t-1)-t}, w_t)\} \subseteq \Gamma_{b2}$ such that $E(w_i, w_j, \alpha) = 1$, where $i = j$; $1 \leq i \leq s \leq 1$, $1 \leq j \leq t$, $s = t$, $s \geq 3$.

*3) non-root relation*: Two BMs form a *non-root* relation if a non-root BU of one BM is *equivalent* to a non-root BU of another BM (Fig. 8).

*Definition 17: non-root* relation: A BM $b_1$ has a *non-root* relation with another BM $b_2$ if $\exists w_p \in \{N_{b1} \cup L_{b1}\}$ and $w_c \in \{N_{b2} \cup L_{b2}\}$ such that for each $(w_p, w_c)$, $E(w_p, w_c, \alpha) = 1$. We denote it by $R_{n-r}(b_1, b_2) = \{(w_p, w_c): w_p \in \{N_{b1} \cup L_{b1}\}, w_c \in \{N_{b2} \cup L_{b2}\}, E(w_p, w_c, \alpha) = 1\}$.

Furthermore, the non-root relation can be of three different types: *branch-branch relation*, *leaf-branch relation* and *leaf-leaf relation*.

*4) branch-branch relation:* If an intermediate BU of a BM is *equivalent* to an intermediate BU of another BM, they form a branch-branch relation (Fig. 8(a)).

*Definition 18: branch-branch* relation: A BM $b_1$ has a *branch-branch relation* with another BM $b_2$ if $\exists w_p \in N_{b1}$ and $w_c \in N_{b2}$ such that for each $(w_p, w_c)$, $E(w_p, w_c, \alpha) = 1$. We denote it by $R_{b-b}(b_1, b_2) = \{(w_p, w_c): w_p \in N_{b1}, w_c \in N_{b2}, E(w_p, w_c, \alpha) = 1\}$.

*5) leaf-branch relation:* If a leaf BU of a BM is *equivalent* to a branch node of another BM, they form a leaf-branch relation (Fig. 8(b)).

*Definition 19: leaf-branch* relation: A BM $b_1$ has a *leaf-branch relation* with another BM $b_2$ if $\exists w_p \in L_{b1}$ and $w_c \in N_{b2}$ such that for each $(w_p, w_c)$, $E(w_p, w_c, \alpha) = 1$. We denote it by $R_{l-b}(b_1, b_2) = \{(w_p, w_c): w_p \in L_{b1}, w_c \in N_{b2}, E(w_p, w_c, \alpha) = 1\}$.

*6) leaf-leaf relation:* If a leaf BU of a BM is *equivalent* to a leaf BU of another BM, they form a leaf-leaf relation (Fig. 8(c)).

*Definition 20: leaf-leaf* relation: A BM $b_1$ has a *leaf-leaf relation* with another BM $b_2$ if $\exists w_p \in L_{b1}$ and $w_c \in L_{b2}$ such that for each $(w_p, w_c)$, $E(w_p, w_c, \alpha) = 1$. We denote it by $R_{l-l}(b_1, b_2) = \{(w_p, w_c): w_p \in L_{b1}, w_c \in L_{b2}, E(w_p, w_c, \alpha) = 1\}$.

## V. MAPPING INTEGRATION RELATIONS TO DEFEFCTS DETECTION

An SRS may have many different types of defects. In this section, we map the previously defined integration relations to some particular types of requirements defects that may exist in the SRS. We identify four primary types of potential defects in the requirements: *incomplete*, *ambiguous*, *incorrect* and *redundant*. Each of these potential defects can be detected *automatically*, and an issue can be documented upon the verification performed by the requirements analyst. We demonstrate the mappings below:

*A. Incomplete*

If the root BU of a BM is not *equivalent* to any BU of all other BMs, the precondition BU of the former BM is out of context and the requirements associated with it are *incomplete*. Therefore, a BM (and its associated requirement in the SRS) is *incomplete* if it does not have any integration relation unless otherwise accepted as an *initialisation* BM by the requirements analyst. A BM that sets up the initial states of the software system is referred as the *initialisation* BM.

*Definition 21:* An initialisation BM, denoted by $b_{init}$, sets up the initial states of the software system.

*Definition 22:* A BM $b_i$ is *incomplete* if it does not have any primary types of integration relation i.e. $\forall b_j, i \neq j, R(b_j, b_i) = \Phi$, $b_i \neq b_{init}$.

### B. Ambiguous

There might be more than one interpretation of the requirements and their relations if two BMs form *multi-preconditions* relation. If the requirements analyst does not accept all the parent integration BUs, it implies that requirements associated with them are *ambiguous* since there may exist multiple interpretations. The most obvious reason is that the child integration BU is a weak precondition that needs to be modified to avoid forming *multi-preconditions* relation. If this is not a suitable option, a new precondition BU must be added to resolve this ambiguity.

*Definition 23:* If a *multi-preconditions* relation, formed by two BMs $b_1$ and $b_2$ i.e. $R_{m-p}(b_1, b_2) = 1$, is not accepted by the requirements analyst, the requirements associated with $b_1$ and $b_2$ are ambiguous.

### C. Incorrect

If there exists one or more *non-root* relations and no other primary integration relations between two BMs, the preconditions of these BMs may have been described inaccurately. Therefore, the BMs forming the *non-root* relations are incorrect. If the requirements analyst accepts this relation, an issue of inaccurately specified requirements can be documented automatically.

*Definition 24:* The pair of BMs $b_1$ and $b_2$ are incorrect if they form *non-root* relation only i.e. $R_{n-r}(b_1, b_2) \neq \Phi$ and $R(b_1, b_2) = \Phi$.

### D. Redundant

In the SRS, some requirements may be *redundant*. Redundancy itself is not an error, but it can easily lead to errors in a later stage of development. If two BMs form a *sub-path* relation, they represent redundant information. This redundancy can be identified and documented automatically if the requirements analyst accepts the sub-path relation.

*Definition 25:* The pair of BMs $b_1$ and $b_2$ contains *redundant* requirements if they form a *sub-path* relation i.e. $R_{s-p}(b_1, b_2) = 1$.

In the following section, we elaborately describe the mappings of defects detection from the integration relations with a case study.

## VI. CASE STUDY: MICRO-WAVE OVEN SYSTEM

In this section, we use a micro-wave oven system case study to demonstrate defects detection by formal integration relations. The requirements of a micro-wave oven system are given in Table III. We used the requirements of the micro-wave oven as published in [1] and subsequently modified them by injecting defects to demonstrate our approach.

Fig. 9 shows the BMs (represented as BTs) and their integration relations of the Micro-wave oven system. From the requirements R1-R9, we get $b_1$ to $b_{10}$ BMs. Here, we considered the same settings used to illustrate Definition 10.

### A. Results

The BMs form different primary and special types of integration relations as defined in the Section IV. From these integration relations, Table I enlists potential defects according to the mappings defined at Section V.

**Table I: Defects Detection by Integration Relations**

| BM | Integration Relation | Defects Type | Issue |
|---|---|---|---|
| $b_8$ | no relation | Incomplete | The precondition of $b_8$ is missing. That means, it is unknown when this behaviors occur. |
| $b_9$, $b_2$ | multi-preconditions relation | Ambiguous | If this relation is unaccepted, the requirements associated with them are ambiguous and there exist multiple interpretation of integration. |
| $b_{10}$, $b_8$ | non-root relation | Incorrect | The requirements associated with $b_8$ and $b_{10}$ have been described inaccurately. |
| $b_1$, $b_9$, $b_6$ | sub-path relations | Redundant | Same sequence of behaviors have been mentioned in the requirements associated with the BMs. |

The BM $b_8$ has no relation with other BMs, hence $b_8$ is an *incomplete* requirement. If the *multi-preconditions* relation of $b_9$ and $b_2$ is not accepted by the requirements analyst, they are characterised by an *ambiguous* interpretation of the associated requirements for integration. The *non-root* relation in $b_{10}$ and $b_8$ implies that these requirements are *incorrect*. Finally, the BMs $b_6$, $b_9$ and $b_1$, $b_9$ form *sub-path* relations which imply *redundant* requirements specified in the SRS.

Besides these special relations, $b_7$ and $b_4$ form *leaf-root* relation; $b_7$ forms *root-root* relations both with $b_3$ and $b_5$, both of which have *branch-root* relations with $b_6$ (not shown in Fig. 9); $b_6$ and $b_7$ form *branch-root* and *leaf-root* relations respectively.

### B. Comparative Study

Defects detection techniques are broadly divided into two approaches, *non-systematic* and *systematic*. Many studies have been performed to compare these two approaches. Laitenberger et al. [5] provided evidence that *perspective-based reading* (PBR), a systematic approach, is more effective in UML diagrams than *non-systematic ad-hoc* and *check-list* based approaches. Contrary to this, Lanubile and Visaggio [18] evaluated PBR techniques having three different perspectives (Use Case Analyst, Structured Analyst, Object-Oriented Analyst) with *non-systematic* approaches, and they did not find any difference between *systematic* and *non-systematic* approaches with respect to the percentage of discovered defects.

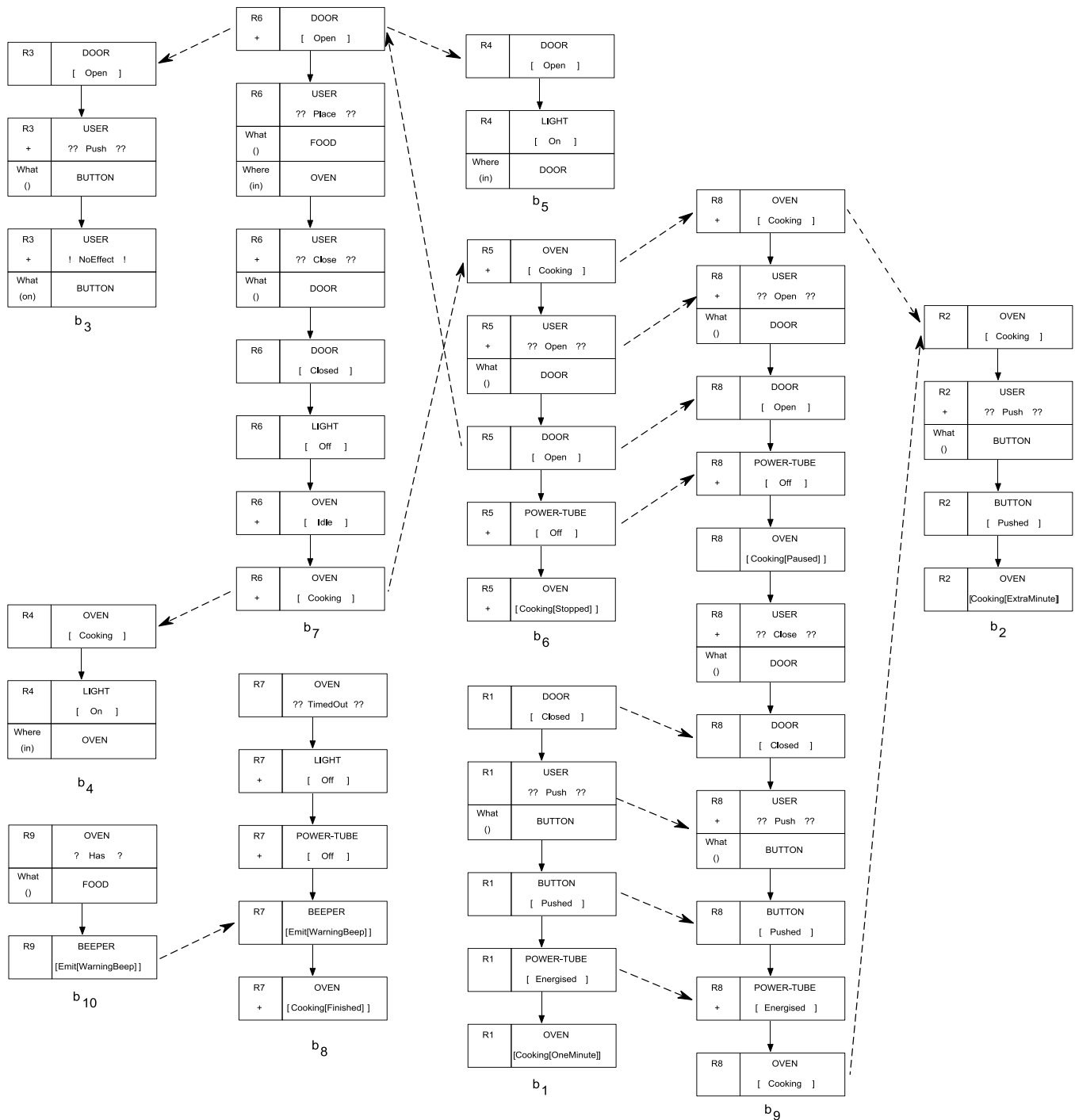

Fig. 9. Integration Relations in BMs of Micro-wave Oven

Both these results can be contrasted with Dromey and Powell [6], who conducted a comparison of finding defects by applying BE and traditional review techniques with an industrial case study. Their study showed that BE (35 defects) has a double rate of defect detection over conventional inspections (17 defects). They showed that although the defect discovery rate is higher for the traditional review, the lower analysis rate of BE resulted in a significantly higher detection of severe defects.

Since BE is currently performed as a manual and ad-hoc process, our semi-automated techniques are more likely to increase the defects discovery rate in the integration stage of BTs. Moreover, our proposed techniques find two additional types of defects (*ambiguous* and *incorrect*) in that integration stage. As a result, our proposed approach makes BE superior to the traditional review. Moreover, BE results in a specification that can be validated and verified prior to software system development, which is *unique* in comparison to all other techniques of RE. Table II shows the summary of the improvements we made over BE.

**Table II: Improvements on BE having an underlying framework.**

| No | Task | BE | BE with underlying framework |
|---|---|---|---|
| 1 | Discovery of integration points | manual, ad-hoc | semi-automated, formal |
| 2 | Identification of integration relations | manual, ad-hoc | semi-automated, formal |
| 3 | Special relations | no study | formal study |
| 4 | Defects Detection Techniques | manual, ad-hoc | semi-automated, formal |
| 5 | Types of defects detected at integration | *Incomplete, redundant* | *incomplete, ambiguous, incorrect, redundant* |

*C. Limitations*

Since defects detection, to a large extent, depends on domain information, our approach has the following limitations:

1) Requirements analyst has to determine the integration strategy and confirm any relations. They also have to confirm any defect that is detected, excluding *incompleteness*.

2) Defects may remain present in the specification even after the application of this approach.

## VII. RELATED WORK

We classify the related work into three categories:

*A. Formalisation of BE*

The BMP allows representing potentially vastly different types of requirements into BTs and merging them into a formal specification BT. The semantics of the specification BT has been provided by Colvin and Hayes using an extension of Communicating Sequential Process (CSP) [11]. However, the formalisation of the BMP for deriving the specification BT from the NL requirements is still missing.

The integration of BTs is based on two informal axioms namely *precondition* and *interaction* [1]. Winter et al. [19] broadly discussed an overall approach to BTs integration with a framework of rules using a notation-independent graphical model. However, they do not provide a detailed mathematical framework to formalise the graphical form of BTs. Our approach also differs in defining different types of primary and special integration relations, which are necessary to formalise the integration strategy of the BMP and support our defects detection approach.

*B. Defects Detection*

Many different defect detection techniques [20] have been proposed so far. Negotiation [21], Inquiry [22] and Inspections ([4], [23], [24]) are human-centred requirement analysis techniques where stakeholders' inputs are utilised to analyse the correctness of the models.

Among these, Inspections have proved to be an effective means of defect detection. *Ad-hoc* and *check-list* based approaches seem to be *primitive*, where *scenario*, *perspective* and *usage* based approaches are more *systematic*. There also exists a family of defect detection techniques for UML diagrams ([3], [4]). Many comparative studies ([5], [18], [25]) indicate that all these techniques are effective, and sometimes one technique outperforms another in different experimental settings.

BE provides a systematic way to analyse the quality of the SRS, and discover defects in the requirements while modeling the requirements. BE has been proved to be more effective that the traditional review [6]. However, the whole process along with the defect detection is informal and on an ad-hoc basis, which is what we have partially addressed in this paper.

*C. Applications of BE*

BE has been extended in many ways in the field of RE. Wen and Dromey ([26]) proposed an informal traceability model to determine the impact of a system change on the architecture and design of the system by embedding the changes in the specification BT. Wendland et al. [27] developed a requirements model by augmenting test related information into BT to derive a test specification of a software system. A number of tools ([28]–[30]) have been developed to support BE.

Myers et al. [12] developed a co-modeling approach to simulate the behavioral scenarios of different set of software/hardware partitions into Modelica framework [31], an equation-based, object-oriented mathematical modeling language to model complex physical systems in multiple domains.

BT has been used for model-checking to assess the safety requirements [13], analyse role-based access control [14], perform Failure Modes and Effects Analysis (FMEA) [15] for determining hazardous states, to identify the combinations of component failures by Cut Set Analysis (CSA) [16], verify large systems using a slicing techniques [17] and so on.

## VIII. CONCLUSION AND FUTURE WORK

In this paper, we provided a formal mathematical framework for BTs and their different types of integration relations. This framework formalises an important part of the integration process of the BMP in greater details. We also presented semi-automated defects detection techniques utilising that framework. While our defects detection techniques do not guarantee to find all the defects in the SRS, it still is useful for detecting substantial amount of requirements defects. In addition, this foundational work opens many more future directions to undertake:

1) *Formalisation of integration process of BTs:* Formalising a semi-automated integration process of BTs using the mathematical framework is our current research focus.

2) *Formalisation of specification process of BTs:* Our future research focus is to formalise a semi-automated specification process by extending the framework.

3) *Change Management Framework:* Another key motivation of our work is that this framework creates the foundation of a change management framework for requirements modeling, which is another focus of future research.

4) *Round-trip Engineering:* Our framework may also be the basis of round-trip engineering to automatically propagate the effect of a change in one software artefact backward or forward to other software artefacts at different stages of the process.

## APPENDIX A

**Table III: Requirements of the Micro-wave Oven System**

| R1 | There is a single control button available for the use of the oven. If the oven door is closed and you push the button, the oven will start cooking (that is, energise the power-tube) for one minute. |
|---|---|
| R2 | If the button is pushed while the oven is cooking, it will cause the oven to cook for an extra minute. |
| R3 | Pushing the button when the door is open has no effect. |
| R4 | Whenever the oven is cooking or the door is open, the light in the oven will be on. |
| R5 | Opening the door stops the cooking. |
| R6 | Closing the door turns off the light. This is the normal idle state, prior to cooking when the user has placed food in the oven. |
| R7 | If the oven times out, the light and the power-tube are turned of and then a beeper emits a warning beep to indicate that the cooking has finished. |
| R8 | While the oven is cooking, if the user opens the door, the oven pauses the cooking and resumes if the door is closed and user pushes the button. |
| R9 | If the food is inside the oven, the beeper beeps. |